\documentclass[prl,twocolumn,showpacs]{revtex4}
\usepackage{graphicx}
\usepackage{color}

\newcommand{\pa}{\partial}

\begin{document}
\title{Remark on ``Pair Creation Constrains Superluminal Neutrino Propagation"}

\author{Sen Hu$^{1,2}$\footnote{shu@ustc.edu.cn}, Wei Huang$^{1,3}$\footnote{weihuang@mail.ustc.edu.cn}, Si Li$^4$\footnote{sili@math.northwestern.edu}, Mu-Lin Yan$^{1,3}$\footnote{Corresponding author. mlyan@ustc.edu.cn}}
\affiliation{$^1$Wu Wen-Tsun Key Lab of Mathematics of Chinese Academy of Sciences\\
$^2$School of Mathematical Sciences,\\
$^3$Department of Modern Physics,\\
University of Science and Technology of China, Hefei, Anhui 230026, China\\
$^4$Mathematics Department, Northwestern University, Evanston, IL 60201
}

\begin{abstract}
The concept of group velocity of a particle should be consistent with its Hamilton-Jacobi velocity. This point is missed in the work of Cohen, Glashow, ``{\it Pair Creation Constrains Superluminal Neutrino Propagation}" (Phys. Rev. Lett. {\bf 107}, 181803 (2011)). It then leads to the conclusion of existence of Cherenkov-like radiation provided one sees superluminal neutrinos. We show that in the framework of Special Relativity with de Sitter space-time symmetry (dS-SR) the above Cohen-Glashow argument does not hold and the Cherenkov-like radiation is forbidden.
\end{abstract}



\maketitle

\vskip0.3in
In \cite{CG}, the authors show that superluminal neutrinos may lose energy rapidly via the bremsstrahlung of electron-positron
pairs ($\nu\rightarrow \nu+e^-+e^+$) (i.e., the Cherenkov-like radiation in vacuum). Given the claimed superluminal velocity and at the stated mean energy in the OPERA report \cite{OPERA}, the authors of Ref. \cite{CG} find
that most of the neutrinos would have suffered several pair emissions en route, causing the beam to be
depleted of higher energy neutrinos. ``{\it This presents a significant challenge to the superluminal interpretation of the OPERA data."} \cite{CG}.

However, the conclusion in \cite{CG} relies on two {\it ad hoc} assumptions: 1. the neutrino's group velocity $v_\nu=dE_\nu/dp_\nu$, and 2. the dispersion relation violating Lorentz symmetry for $\nu$ is $E_\nu=\sqrt{(1+\delta)}c p_\nu$ (we do not take usual convention of $c=1$ for clarity). We would like to address in this Letter that the conclusion fails to be true in general, and hence cannot be regarded as the criterion determining the superluminary of particles broken Lorentz symmetry. The remarks are as follows:

\begin{enumerate}
\item  In the experiment a neutrino is treated as a particle with a small mass. The group velocity represents the propagating velocity of a particle's energy-momentum. From the mechanics principle, the group velocity $v_\nu\equiv \dot{x}(\nu)$ is determined by means of the canonical Hamiltonian equations:
\begin{eqnarray}\label{1}
&&\dot{x}(\nu)={\pa H(\nu) \over \pa \pi(\nu)},   \\
\label{2} &&\dot{\pi}(\nu)=-{\pa H(\nu) \over \pa x},
\end{eqnarray}
where $H(\nu)=H(t,x,\pi)\equiv \dot{x}\pi-L(t,x,\dot{x})$ and $\pi(\nu)=\pa L(t,x,\dot{x})/\pa \dot{x}$ are the system's Hamilton (or canonical energy) and the canonical momentum of $\nu$ respectively, and hence $v_\nu\equiv \dot{x}(\nu)$ means that the group velocity is the Hamiltonian-Jacobi velocity of the system (see Eq. (\ref{1})).

In order to see the connection more explicitly between the group velocity and the Hamiltonian-Jacobi velocity mentioned above, let us examine it in the framework of Einstein Special Relativity (E-SR). E-SR has Poincar\'e-Lorentz space-time symmetry, and the action is $S_{\rm E-SR}=-m_\nu c\int \sqrt{\eta_{\mu\nu}dx^\mu dx^\nu}\equiv \int dt L_{\rm E-SR}(\nu)$ with $\eta_{\mu\nu}=diag\{+,-,-,-\}$. Hence we have
\begin{eqnarray}\label{3}
L_{\rm E-SR}(\nu)=-m_\nu c^2 \sqrt{1-{v^2\over c^2}},
\end{eqnarray}
where $v=\dot{x}$ (for simplicity, the component-superscript $i$ were omitted). Then the $H(\nu)$ and $\pi(\nu)$ read
\begin{eqnarray}\label{4}
&&H(\nu)=\sqrt{\pi(\nu)^2c^2+m_\nu^2c^4},   \\
\label{5} &&\pi(\nu)={m_\nu v\over \sqrt{1-v^2/c^2}}.
\end{eqnarray}
The point here is that the $L_{\rm E-SR}(\nu)$ (and $H(\nu)$) is time free and coordinate free.
Comparing Eqs. (\ref{4}) (\ref{5}) with the Noether charges of E-SR's space-time symmetry (i.e., the Poincar\'e-Lorentz symmetry) $p_\nu=m_\nu v\gamma={m_\nu v\over \sqrt{1-v^2/c^2}}$ and $E_\nu=m_\nu c^2\gamma$ (e.g., see {\it pp581-586} and {\it Part 9} in Ref. \cite{Noether}), we have
\begin{eqnarray}\label{6}
E_\nu=H(\nu),~~~~p_\nu=\pi(\nu).
\end{eqnarray}
The usual group velocity formula $v_\nu\equiv \dot{x}=\pa H(\nu)/\pa \pi(\nu)=dE_\nu/dp_\nu$ has been proved for the time- and coordinate-independent Hamiltonian system by means of the Hamiltonian-Jacobi formalism. Obviously, suppose $L_{\rm E-SR}(\nu)$ (and $H(\nu)$) was not time free and coordinate free, Eq. (\ref{6}) would not be true and then $v_\nu\neq dE_\nu/dp_\nu$.  We would like to address here that the space-time symmetry of SR in general could be de Sitter symmetry, which is larger than Poincar\'e-Lorentz symmetry \cite{look,Lu74,Ours} (i.e., (algebra of Poincar\'e group)=(algebra of de Sitter group)$|_{R\rightarrow \infty}$ \cite{Ours}) . As is well known that the symmetries are crucial to construct SR theory. There is no {\it a priori} reason to assume the space-time symmetry is Poicar\'e-Lorentz or de Sitter. For the de Sitter case, $H(\nu)$ is time- and coordinate-dependent,  while the energy-momenta are conserved due to the symmetry of space-time (see below).
Therefore, careful consideration of $v_\nu$ is necessary.

In short, we have shown that the general group velocity formula is Eq. (\ref{1}): $v_\nu\equiv \dot{x}=\pa H(\nu)/\pa \pi(\nu)$ instead of $v_\nu=dE_\nu/dp_\nu$. The later is only the degenerate form of Eq. (\ref{1}). Namely $v_\nu=dE_\nu/dp_\nu$ is true only for the time-free and coordinate-free Hamiltonian system. This is a strong condition. However this condition is not considered in \cite{CG}, and hence the applicability of the conclusion of \cite{CG} is limited. Or, that conclusion fails to be true in general.

\item The model's dispersion relation $E_\nu=\sqrt{(1+\delta)}c p_\nu$ was introduced in \cite{CG} for discussion of particle's superluminal phenomenons. Besides the relation is artificial, the mechanism to yield superluminary  via $v=dE_\nu/dp_\nu$ is also artificial.

\item In order to show the meaning of above remarks more concretely, let us consider the Special Relativity with de Sitter space-time symmetry (dS-SR) \cite{look, Lu74}. The Hamiltonian formalism of dS-SR has been formulated in \cite{Ours}, which is based on Beltrami metric $B_{\mu\nu}(x)=\frac{\eta_{\mu\nu}}{\sigma (x)}
+\frac{\eta_{\mu\lambda}\eta_{\nu\rho} x^\lambda x^\rho}{R^2 \sigma(x)^2}$ with $\sigma(x) \equiv  1-\frac{1}{R^2}\eta_{\mu\nu}x^\mu x^\nu$.
The action and Lagrangian read
\begin{eqnarray} \nonumber
&& S\equiv\int dt L_{\rm dS-SR}(t,x^i,\dot{x}^i)=-m_\nu c\int dt {\sqrt{B_{\mu\nu}(x)dx^\mu dx^\nu} \over dt}\\
\label{B1}&&L_{\rm dS-SR}(t,x^i,\dot{x}^i)=-m_\nu c\sqrt{B_{\mu\nu}(x)\dot{x}^\mu \dot{x}^\nu}
\end{eqnarray}
From the mechanics principle, the 10 conserved Noether charges in dS-SR  are as follows: \cite{Ours}
\begin{eqnarray}\nonumber
p^i &=& m_\nu \Gamma \dot{x}^i
\\
\nonumber  E &=&  m_\nu c^2 \Gamma
\\
\label{physical momenta} K^{i} & =& m_\nu c \Gamma (x^i -t\dot{x}^i)=m_\nu c\Gamma x^i-tp^i
\\
\nonumber L^{i} & =& -m_\nu \Gamma \epsilon^{i}_{\;jk} x^j
\dot{x}^k=-\epsilon^{i}_{\;jk} x^jp^k.
\end{eqnarray}
\noindent Here $E,{\mathbf p},{\mathbf L},{\mathbf K}$ are
conserved physical energy, momentum, angular-momentum and boost
charges respectively, and $ \Gamma $ is:
\begin{widetext}
 {\begin{eqnarray} \label{new parameter}
 \Gamma^{-1}\hskip-0.1in =\sigma(x) \frac{ds}{c dt}=\frac{1}{R} \sqrt{(R^2-\eta_{ij}x^i
x^j)(1+\frac{\eta_{ij}\dot{x}^i \dot{x}^j}{c^2})+2t \eta_{ij}x^i
\dot{x}^j -\eta_{ij}\dot{x}^i \dot{x}^j t^2+\frac{(\eta_{ij}
x^i\dot{x}^j)^2}{c^2}}.
\end{eqnarray}}
\end{widetext}
The dispersion relation reads \cite{Ours}
\begin{eqnarray}\label{dp}
E^2 =m_\nu^2 c^4+{\mathbf p}^2 c^2 + \frac{c^2}{R^2}
({\mathbf L}^2-{\mathbf K}^2).
\end{eqnarray}
The canonical momentum $\pi_{i} =\frac{\pa L_{\rm dS-SR}}{\pa
\dot{x}^i} = -m_\nu \sigma(x) \Gamma B_{i \mu}\dot{x}^{\mu}\neq p^i$, and the canonical energy (or Hamiltonian) $H
=\sum_{i=1}^3 \frac{\pa L_{\rm dS-SR}}{\pa \dot{x}^i} \dot{x}^i
-L_{\rm dS-SR}=m_\nu c \sigma(x) \Gamma B_{0 \mu}\dot{x}^{\mu}\neq E$ (see Eq. (\ref{physical momenta})). When the neutrino ($\nu$-) kinematics was discussed in the framework of dS-SR, above consequences lead to the following equations which are against Eq. (\ref{6}):
\begin{eqnarray}\label{6-1}
E_\nu\neq H(\nu),~~~~p_\nu\neq \pi(\nu),
\end{eqnarray}
where the super-(or sub-)scripts of $p$ and $\pi$ are emitted. Therefore neutrino's
velocity $v_\nu\equiv \dot{x}=\pa H(\nu)/\pa \pi(\nu)\neq dE_\nu/dp_\nu$. Namely, we have explicitly shown that when  $L(\nu)$ (and $H(\nu)$) is not time free and coordinate free, the usual formula $v_\nu= dE_\nu/dp_\nu$ does not hold and it cannot be used.

The dS-SR kinematics of neutrinos $\nu$ in the OPERA experiment has been discussed in Ref.\cite{Ours1}. Taking the origin of the reference space-time frame to be the Big Bang (BB) occurred point, the OPERA neutrino's coordinates and time are: $|x^i/R|\simeq 2\times 10^{-23}$ and $ct/R\simeq 6.7\times 10^{-3}$ (where $R\simeq 2\times 10^{12} l.y.$ has been used \cite{Ours1}). Then we have
\begin{eqnarray}\nonumber
 &&v_{\nu}={\pa H(\nu)\over \pa \pi(\nu)}\simeq c \sqrt{1-\frac{m_\nu^2 c^4}{E_{\nu}^2}\over 1-\frac{c^2 t^2}{R^2}}\simeq c(1+{c^2t^2\over 2R^2})\\
\label{19} &&\hskip0.4in \simeq (1+2.4\times 10^{-5}) c >c.
\end{eqnarray}
Therefore from the prediction of dS-SR kinematics, the OPERA neutrinos are superluminal.
Next, we examine whether the process of Cherenkov-like radiation in vacuum $\nu(p)\rightarrow \nu(p')+e^-(k')+e^+(k)$ associating with the superluminal neutrinos $\nu(p)$ takes place or not. To OPERA neutrinos, Eq. (\ref{dp}) becomes dispersion relation violating Lorentz symmetry which is as follows
\begin{equation}\label{CG4}
p^2c^2={E^2-m_\nu^2c^4\over 1-{c^2t^2\over R^2}}.
\end{equation}
Substituting Eq. (\ref{CG4}) into the threshold equation in \cite{CG} \cite{mli} $(E^2-p^2c^2)_{\rm thr.}=(2m_e+m_\nu)^2c^4$, we get
\begin{eqnarray}\nonumber
&&E_{\rm thr.}^2=-{R^2\over c^2t^2}[(2m_e+m_\nu)^2c^4(1-{c^2t^2\over R^2})
-m_\nu^2c^4]\\
\label{CG5} && \hskip0.3in \simeq -{R^2\over c^2t^2}4m_e^2c^4.
\end{eqnarray}
 Obviously, there is no real and positive solution of $E_{\rm thr.}$ from Eq. (\ref{CG5}) for $R^2>0$ . This fact indicates that the Cherenkov-like process $\nu(p)\rightarrow \nu(p')+e^+(k)+e^-(k')$ claimed by Ref.\cite{CG} for superluminal $\nu(p)$ with $v_\nu >c$ (see Eq. (\ref{19})) is forbidden kinematically.

\item From Eq.(\ref{CG5}), the conclusion that there is no physical solution of $E_{\rm thr.}$ is irrelevant to the magnitude of $R$, or the data of OPERA \cite{OPERA}. Namely, regardless of what is the final results of OPERA experiment in the future, Eq. (\ref{CG5}) has no real, positive and finite solution of $E_{\rm thr.}$, and hence the process of
($\nu\rightarrow \nu+e^-+e^+$) is always forbidden.

\item According to above, it has been learned that the null experiments to observe the energy loss via $\nu\rightarrow \nu+e^++e^-$ cannot be used to rule out the possibility of existence of superluminal neutrinos. For example, the null results of ICARUS \cite{ICARUS} experiment cannot lead to consequences without superluminal neutrinos. Moreover, besides our model in \cite{Ours1},  the phenomenological models based on other considerations with superluminal neutrinos and without energy-loss via the Cherenkov-like radiation claimed by \cite{CG} have also been constructed and proposed, e.g. see \cite{Ellis} \cite{ChangZ}. This indicates also that the conclusion of \cite{CG} fails to be generic.

\end{enumerate}
\vskip0.1in
\noindent {\bf Conclusion:} The concept of group velocity of a particle should be consistent with its Hamilton-Jacobi velocity. The usual definition of $v_\nu=dE_\nu/dp_\nu$ is only true for the time-free and coordinate-free Hamiltonian system. Otherwise, $v_\nu=\pa H_\nu/\pa \pi_\nu$ represents the
particle's velocity. This point is not considered in Ref.\cite{CG}. It then leads to the conclusion of ``{\it superluminal neutrinos may lose energy rapidly via the bremsstrahlung of electron-positron pairs ($\nu\rightarrow \nu+e^-+e^+$)}" claimed in \cite{CG} is not always valid. When we work with the SR with de Sitter space-time symmetry (dS-SR) we have to deal with time- and coordinate-dependent Hamiltonian system. It has been proved that when neutrino's Hamilton-Jacobi velocity is superluminal in the dS-SR kinematics, the process of ``{\it bremsstrahlung of electron-positron pairs ($\nu\rightarrow \nu+e^-+e^+$)}" claimed by \cite{CG} is forbidden. 

\begin{center}{\bf ACKNOWLEDGMENTS}
\end{center}
{ This work is partially supported by
National Natural Science Foundation of China under Grant No.~10975128
and No.~11031005 and by the Wu Wen-Tsun Key Laboratory of Mathematics at USTC of Chinese Academy of Sciences.}



\end{document}